\documentclass[prl,aps,reprint,superscriptaddress,showpacs]{revtex4-2}
\usepackage{amssymb}
\usepackage{amsmath}
\usepackage{dcolumn}
\usepackage{bm}
\usepackage{footmisc}
\usepackage{graphicx}
\usepackage{subfigure}
\usepackage{mathrsfs}
\usepackage{autobreak}
\usepackage[T1]{fontenc}
\usepackage[utf8]{inputenc}
\usepackage{hyperref}
\usepackage{color, soul}

\allowdisplaybreaks
\begin{document}
\title{Surpassing Quantum Noise Limits with Nonlinear Amplification}

\author{Ya-Long Ren}
\homepage{These authors contributed equally to this work.}
\affiliation{MOE Key Laboratory for Nonequilibrium Synthesis and Modulation of Condensed Matter,
Shaanxi Province Key Laboratory of Quantum Information and Quantum Optoelectronic Devices, School of physics,
Xi'an Jiaotong University, Xi'an 710049, China}
\affiliation{Hefei National Laboratory, Hefei 230088, China}
\author{Rong-Teng Cao}
\homepage{These authors contributed equally to this work.}
\affiliation{MOE Key Laboratory for Nonequilibrium Synthesis and Modulation of Condensed Matter,
Shaanxi Province Key Laboratory of Quantum Information and Quantum Optoelectronic Devices, School of physics,
Xi'an Jiaotong University, Xi'an 710049, China}
\author{Sheng-Li Ma}
\email{msl1987@xjtu.edu.cn}
\affiliation{MOE Key Laboratory for Nonequilibrium Synthesis and Modulation of Condensed Matter,
Shaanxi Province Key Laboratory of Quantum Information and Quantum Optoelectronic Devices, School of physics,
Xi'an Jiaotong University, Xi'an 710049, China}
\author{Ren Zhang}
\affiliation{MOE Key Laboratory for Nonequilibrium Synthesis and Modulation of Condensed Matter,
Shaanxi Province Key Laboratory of Quantum Information and Quantum Optoelectronic Devices, School of physics,
Xi'an Jiaotong University, Xi'an 710049, China}
\affiliation{Hefei National Laboratory, Hefei 230088, China}
\author{Fu-Li Li}
\affiliation{MOE Key Laboratory for Nonequilibrium Synthesis and Modulation of Condensed Matter,
Shaanxi Province Key Laboratory of Quantum Information and Quantum Optoelectronic Devices, School of physics,
Xi'an Jiaotong University, Xi'an 710049, China}
\author{Franco Nori}
\affiliation{RIKEN Center for Quantum Computing, RIKEN, Wakoshi, Saitama, 351-0198, Japan}
\affiliation{Quantum Research Institute and Physics Department, The University of Michigan, Ann Arbor, Michigan 48109-1040, USA}
\author{Peng-Bo Li}
\email{lipengbo@mail.xjtu.edu.cn}
\affiliation{MOE Key Laboratory for Nonequilibrium Synthesis and Modulation of Condensed Matter,
Shaanxi Province Key Laboratory of Quantum Information and Quantum Optoelectronic Devices, School of physics,
Xi'an Jiaotong University, Xi'an 710049, China}

\begin{abstract}
Linear quantum amplifiers are indispensable tools for quantum technologies, yet their performance is fundamentally limited by quantum noise, precluding any signal-to-noise ratio (SNR) enhancement unless supplemented by post-selection or non-classical resources. To surpass this limitation, we propose a nonlinear quantum amplification strategy that exploits the interplay between a gain-stabilized bright eigenmode of a coupled two-mode bosonic system and Kerr nonlinearity. We demonstrate that this interplay enables the signal gain to surpass the noise gain in a selected quadrature, leading to a net increase in the SNR beyond the quantum limits of conventional linear amplifiers. Our work thus establishes a novel nonlinear amplification paradigm capable of enhancing the SNR, with promising applications across quantum information processing, quantum communications, and quantum metrology.
\end{abstract}

\date{\today}
\maketitle

\emph{Introduction.—}Quantum amplifiers harness quantum-mechanical principles to amplify weak signals with near-minimal added noise \cite{clerk2010introduction,Nation2012,andrekson2020fiber,aumentado2020superconducting,esposito2021perspective,Qin2024a}. As indispensable building blocks, they underpin numerous applications ranging from high-fidelity readout of superconducting qubits \cite{vepsalainen2020impact,sunada2022fast,bengtsson2024model,swiadek2024enhancing,Qin2024b,hazra2025benchmarking} to long-range quantum communications \cite{zhao2023enhancing,grebel2024bidirectional,fesquet2024demonstration,stolk2024metropolitan,liao2025experimental} and ultrasensitive dark-matter detection \cite{braine2020extended,kim2023near,tang2024first,bai2025dark,quiskamp2025near}. Current research focuses predominantly on linear amplifiers \cite{massel2011microwave,macklin2015near,Naoki2016,malz2018quantum,chia2020phase,ZhangQ2020,renger2021beyond,
cochrane2022parametric,zou2024amplifying,vaartjes2024strong,day2024room,kuznetsov2025ultra,dai2025optimizing,malnou2025travelling}, which are classified as phase-insensitive or phase-sensitive based on their noise properties \cite{haus1962quantum,caves1982quantum}. The former treats both quadratures equally and inevitably adds at least half a quantum of noise in the high-gain regime. The latter, conversely, can noiselessly amplify one quadrature at the necessary deamplification of its conjugate counterpart to comply with the Heisenberg uncertainty principle. Both architectures are therefore fundamentally limited by quantum noise, precluding any net enhancement of the signal-to-noise ratio (SNR).

Since the SNR serves as the paramount figure of merit for quantum amplifiers, a central objective is to realize amplification that surpasses the quantum-limited SNR barrier. Probabilistic schemes based on post-selection offer a viable path to circumvent this barrier \cite{Ralph2009,ferreyrol2010implementation,zavatta2011high,donaldson2015experimental,zhao2017quantum,he2021noiseless,guanzon2022ideal,neset2025experimental}.  Nevertheless, their nondeterministic operation and the inherent gain-success probability tradeoff make them less suitable for applications that demand deterministic outputs.  Alternatively, methods that integrate linear amplifiers with non-classical resources (e.g., squeezing or entanglement) to reduce quantum noise are promising \cite{milburn1987linear,ou1993quantum1,ou1993quantum2,kong2013cancellation,malnou2019squeezed,wang2025observation}, but require demanding quantum state engineering.

In this Letter, we present a nonlinear quantum amplification strategy that surpasses the fundamental quantum noise limits of conventional linear amplifiers. Crucially, our approach operates without the need for post-selection or non-classical resources. We demonstrate a net increase in the SNR through simultaneous signal amplification and noise suppression.

Our strategy is based on a general theoretical framework of two coherently coupled bosonic modes, incorporated with gain and Kerr nonlinearity. The underlying mechanism exploits gain to engineer one of the system’s collective eigenmodes with a vanishing effective decay rate. Its interplay with the Kerr nonlinearity then enables a strongly gain-assisted, phase-sensitive amplification; namely, {\itshape the signal gain can exceed the noise gain in a selected quadrature}. Our work thus establishes a new paradigm for enhancing SNR through nonlinear quantum amplification.

\begin{figure}
\centerline {\includegraphics[width=8.5cm]{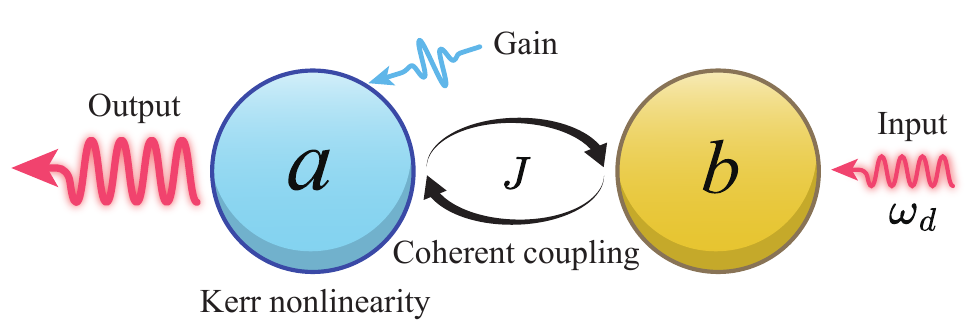}}
\caption{Schematic of the proposed nonlinear quantum amplifier based on a coupled two-mode bosonic system,
where gain and Kerr nonlinearity are incorporated into the mode $a$ (output),
and mode $b$ (input) is driven at a frequency $\omega_d$.}
\end{figure}

\emph{Model.—}As schematically illustrated by Fig.~1, we consider a bipartite quantum system comprising a Kerr-nonlinear bosonic mode $a$ coherently coupled to a linear mode $b$. The system Hamiltonian is given by (we set $\hbar=1$ hereafter)
\begin{align}
H=\omega_{a}a^{\dagger}a+\omega_{b}b^{\dagger}b+Ka^{\dagger2}a^{2}+J(a^{\dagger}b+ab^{\dagger}),
\end{align}
where $\alpha$ ($\alpha=a, b$) is the annihilation operator for mode $\alpha$ with resonance frequency $\omega_{\alpha}$, obeying the commutation relation $[\alpha,\alpha^\dagger]=1$. The parameter $K$ quantifies the Kerr nonlinearity in mode $a$, while $J$ describes the coupling rate between the modes. Here, we focus on the weakly nonlinear regime where $K$ is much smaller than the decay rates $\kappa_{a}$ and $\kappa_{b}$ of each mode. Although this regime naturally occurs in typical experimental platforms, such as optical and microwave cavities, the system’s dissipative nature rules out any amplification effects.

To achieve amplification, we introduce gain to compensate for system losses, creating an eigenmode with vanishing effective decay. When an input field resonantly drives this mode, the resulting excitation is saturated by the Kerr nonlinearity into a stable regime. This nonlinear dynamics give rise to two synergistic results. First, it enables strong nonlinear signal amplification, powered by the gain. Second, and more remarkably, it ensures effective noise suppression in a phase-sensitive way, despite the additional quantum noise inherently introduced by the gain process. This mechanism represents the key finding of this Letter.

Having outlined the physical picture, we now model the full nonlinear amplification dynamics via quantum Langevin equations (QLEs). The model incorporates gain into mode $a$ at a rate $\kappa_g$, while mode $b$ is driven by an input field at the frequency $\omega_d$ (reference phase $\theta_0$). Transforming to a frame rotating at $\omega_d$, we obtain the following QLEs
\begin{align}
\frac{d}{dt}
\left(\!\begin{array}{c}
a\\
b
\end{array}\!\right)
=&-i(\mathcal{H}+2Ka^{\dagger}a\mathcal{M})
\left(\!\begin{array}{c}
a\\
b
\end{array}\right)
+\left(\!\begin{array}{c}
0 \\
\sqrt{2\kappa_{b}}\varepsilon_{\mathrm{in}}
\end{array}\!\right)\nonumber \\
&+\left(\!\begin{array}{c}
\sqrt{2\kappa_{a}}{a}_{\mathrm{in}}+\sqrt{2\kappa_{g}}{g}_{\mathrm{in}}^{\dagger}\\
\sqrt{2\kappa_{b}}{b}_{\mathrm{in}}
\end{array}\!\right)
\end{align}
with the matrices
\[
\mathcal{H} =
\begin{pmatrix}
\Delta_a+i\kappa_{g}-i\kappa_{a} & J \\
J & \Delta_b - i\kappa_b
\end{pmatrix},
\quad
\mathcal{M} =
\begin{pmatrix}
1 & 0 \\
0 & 0
\end{pmatrix},
\]
and the detuning $\Delta_{\alpha}=\omega_{\alpha}-\omega_{d}$.
Here, $\varepsilon_{\mathrm{in}}=\sqrt{2\kappa_{b}N_{{\mathrm{in}}}}\,e^{i\theta_{0}}$ represents the input field amplitude,
where $N_{\mathrm{in}}$ denotes the input mean excitation number.
It should be noted that the loss channel for operator $\alpha$ is linked to the input noise operator $\alpha_{\mathrm{in}}$,
whereas the gain channel for operator $a$ is associated with the input noise operator $g_{\mathrm{in}}^{\dagger}$ \cite{PhysRevLett.123.180501,PhysRevLett.132.243601}.
These noise operators have zero mean $\langle\alpha_{\mathrm{in}}(t)\rangle=\langle{g}^{\dagger}_{\mathrm{in}}(t)\rangle=0$,
and satisfy correlations $\langle\alpha_{\mathrm{in}}(t)\alpha_{\mathrm{in}}^{\dagger}(t^{\prime})\rangle
=\langle{g}_{\mathrm{in}}(t){g}_{\mathrm{in}}^{\dagger}(t^{\prime})\rangle=\delta(t-t^{\prime})$
under the zero-temperature assumption.

In the absence of both the Kerr nonlinearity and the input field, the system is governed by the effective Hamiltonian $\mathcal{H}$ for the two coupled bosonic modes. The eigenfrequencies of $\mathcal{H}$ are given by \cite{sup_mat}
\nocite{DeJesus1987,Mancini1994,Johansson2012,Johansson2013,Lambert2026}
\begin{equation}
\widetilde{\omega}_{\pm}=\frac{1}{2}\left[\widetilde{\Delta} \pm \sqrt{\widetilde{\Delta}^2 + 4(C_1+i C_2)}\right]
\end{equation}
with the complex detuning $\widetilde{\Delta}=\Delta_{a}+\Delta_{b}-i({\kappa_{a}+\kappa_{b}-\kappa_{g}})$, and the coefficients $C_{1}=J^{2}-\Delta_{a}\Delta_{b}-(\kappa_{g}-\kappa_{a})\kappa_{b}$, $C_{2}=\Delta_{a}\kappa_{b}-\Delta_{b}(\kappa_{g}-\kappa_{a})$.
The corresponding eigenmodes read
\begin{equation}
P_{\pm}=m_{\pm}a\pm m_{\mp}b
\end{equation}
with $m_{\pm}=\sqrt{(\widetilde{\omega}_{+}-\widetilde{\omega}_{-}\pm\widetilde{\delta})/2(\widetilde{\omega}_{+}-\widetilde{\omega}_{-})}$,
and $\widetilde{\delta}=\Delta_{a}-\Delta_{b}+i(\kappa_{g}-\kappa_{a}+\kappa_{b})$.
Their decay rates, $\operatorname{Im}(\widetilde{\omega}_{\pm})$, can be tuned via the introduced gain.

By setting $C_1=C_2=0$, the gain rate $\kappa_{g}$ within $\kappa_{a}<\kappa_{g}<(\kappa_{a}+\kappa_{b})$ satisfies
\begin{equation}
\left[\frac{(\omega_{b}-\omega_{a})^{2}}{(\kappa_{g}-\kappa_{a}-\kappa_{b})^{2}}+1\right](\kappa_{g}-\kappa_{a})\kappa_{b}=J^{2},
\end{equation}
under which the eigenmode $P_{-}$ possesses a purely real eigenfrequency and exhibits no decay
with $\operatorname{Im}(\widetilde{\omega}_{-})=0$, whereas the eigenmode $P_{+}$ is dissipative due to $\operatorname{Im}(\widetilde{\omega}_{+})<0$. Since the eigenmode $P_{-}$ has zero decay rate, the mean value of its amplitude remains unchanged over time without driving (See Supplementary Material \cite{sup_mat}). In this sense, $P_{-}$ is identified as a bright eigenmode.

In the presence of the Kerr nonlinearity and under resonant driving of the bright eigenmode at frequency $\omega_d = [\omega_b(\kappa_g-\kappa_a) - \omega_a\kappa_b] / (\kappa_g-\kappa_a-\kappa_b)$, our scheme achieves phase-sensitive amplification that concurrently amplifies the signal and suppresses its quantum noise, leading to an enhancement of the SNR.

We analyze the proposed nonlinear amplification scheme using the standard linearization procedure \cite{aspelmeyer2014cavity}. Each operator is decomposed into a classical mean value and a quantum fluctuation term (e.g., $\alpha=\langle\alpha\rangle+\delta\!\alpha$). This decomposition separates the QLEs into two distinct sets: one for the classical amplitudes, governing the signal gain, and another for the quantum fluctuations, determining the noise properties.

\emph{Signal amplification.—}Below, we study the signal amplification
by investigating the classical equations of motion for the amplitudes \cite{sup_mat}
\begin{equation}
\frac{d}{dt}
\left(\!\begin{array}{c}
{\langle{a}\rangle}\\
{\langle{b}\rangle}
\end{array}\!\right)
=-i(\mathcal{H}+2K|\langle{a}\rangle|^{2}\mathcal{M})
\left(\!\begin{array}{c}
{\langle{a}\rangle}\\
{\langle{b}\rangle}
\end{array}\!\right)
+\left(\!\begin{array}{c}
0\\
\sqrt{2\kappa_{b}}\varepsilon_{\mathrm{in}}
\end{array}\!\right).
\end{equation}
To demonstrate the amplification, we now derive the relation between the output of mode $a$ and the input of mode $b$. This is accomplished by obtaining the steady-state solution of Eq.~(6) (the stability analysis discussed in the Supplementary Material \cite{sup_mat}), from which the input-output relation is extracted
\begin{align}
\langle{a_{\mathrm{out}}}\rangle=\frac{-i2J\sqrt{\kappa_{a}\kappa_{b}}}{2K(\Delta_{b}-i\kappa_{b})|\langle{a}\rangle|^{2}-C_{1}-iC_{2}}\varepsilon_{\mathrm{in}}
\end{align}
with the internal intensity $|\langle{a}\rangle|^{2}$ satisfying the cubic nonlinear equation
\begin{equation}
\begin{aligned}
&\frac{C_{1}^{2}+C_{2}^{2}}{4\kappa_{b}^{2}J^{2}}|\langle{a}\rangle|^{2}
-\frac{K}{\kappa_{b}^{2}J^{2}}(C_{1}\Delta_{b}-C_{2}\kappa_{b})|\langle{a}\rangle|^{4} \\
&+\frac{K^{2}}{\kappa_{b}^{2}J^{2}}(\kappa_{b}^{2}+\Delta_{b}^{2})|\langle{a}\rangle|^{6}=N_{\mathrm{in}}.
\end{aligned}
\end{equation}

\begin{figure}
\centerline {\includegraphics[width=8cm]{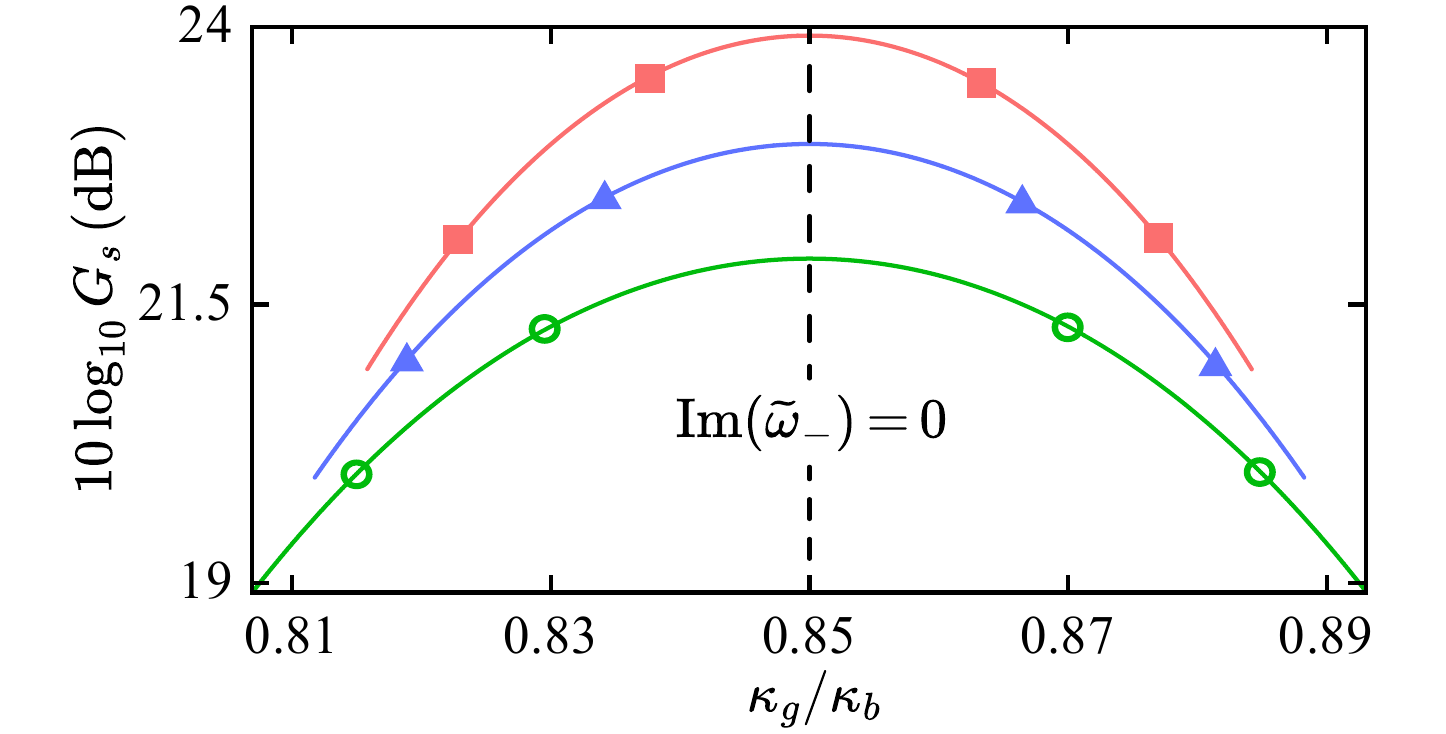}}
\caption{Signal gain $10\mathrm{log_{10}}G_{s}$ versus normalized gain rate $\kappa_g/\kappa_b$ for: $N_\mathrm{in}=0.5$, $K=1\times10^{-4}\kappa_b$ (circle); $N_\mathrm{in}=0.5$, $K=5\times10^{-5}\kappa_b$ (square); $N_\mathrm{in}=0.7$, $K=5\times10^{-5}\kappa_b$ (triangle).
Parameters: $\omega_d=\omega_a-0.3\kappa_b$, $\omega_b=\omega_a+0.2\kappa_b$, $\kappa_a=0.25\kappa_b$, $J=\sqrt{3}\kappa_b/2$, $\theta_s=\mathrm{Arg}(1-0.5i)$.}
\end{figure}

In a purely dissipative system, the output-to-input ratio cannot exceed unity because of intrinsic energy loss. We therefore introduce gain to offset loss and enable amplification. Crucially, when the bright eigenmode emerges (i.e., \(C_1=C_2=0\)), resonant coupling between the input field and this mode leads to continuous energy buildup. This gives rise to divergence in both the internal intensity and the output, implying infinite amplification. The Kerr nonlinearity suppresses this divergence through an intensity-dependent frequency shift of $2K|\langle{a}\rangle|^{2}$, which detunes the system from resonance (see Supplemental Material \cite{sup_mat}). As a result, the internal intensity saturates at a finite value determined by
\begin{align}
|\langle a^{bm} \rangle|^2=\left(\frac{N_{\mathrm{in}}\kappa_{n}\kappa_b}{K^2}\right)^{1/3},
\end{align}
resulting in finite signal amplification. Here, $\kappa_{n}=\kappa_{g}-\kappa_{a}$ is the net gain rate, and the superscript "$bm$" indicates operation at the bright-eigenmode point.

Substituting Eq.~(9) into Eq.~(7), the input-output relation becomes
\begin{align}
\langle{a^{bm}_{\mathrm{out}}}\rangle=\sqrt{\kappa_{a}\!\left(\frac{\kappa_{n}}{\kappa_b^{2}K^2}\right)^{1/3}\!\left(\frac{1}{N_{\mathrm{in}}}\right)^{2/3}}
\varepsilon_{\mathrm{in}}e^{i\theta_s},
\end{align}
where the output acquires a phase shift $\theta_s=\mathrm{Arg}(1-i\Delta_b/\kappa_b)$ relative to the input. Although the Kerr nonlinearity causes the gain saturation effect, the factor $\kappa_{a}({\kappa_{n}}/{\kappa_b^{2}K^2})^{1/3}$ in the regime $K\ll\{\kappa_a,\kappa_b\}$ governs a substantial nonlinear enhancement of the output-to-input ratio, corresponding to strong signal amplification. At the same time, the amplification saturates with increasing $N_{\mathrm{in}}$, as evidenced by the $N_{\mathrm{in}}^{-2/3}$ scaling.

We now quantify the signal amplification via the output quadrature $\langle\mathbb{X}^{a}_{\mathrm{out}}\rangle=({\langle{a}_{\mathrm{out}}\rangle}^{*}e^{i\theta}+{\langle{a}_{\mathrm{out}}\rangle}e^{-i\theta})/\sqrt{2}$, where $\theta=\theta_{s}+\theta_0$ is the total quadrature angle. The corresponding signal gain $G_s$ is defined as
\begin{equation}
G_s=\frac{I_{\mathrm{out}}}{I_{\mathrm{in}}},
\end{equation}
where $I_{\mathrm{out}}=\langle\mathbb{X}^{a}_{\mathrm{out}}\rangle^2$ and $I_{\mathrm{in}}=2|\varepsilon_{\mathrm{in}}|^2$ are the output and input quadrature intensities, respectively. Figure~2 displays the logarithmic signal gain $10\mathrm{log_{10}}G_{s}$ (in dB units) versus the normalized gain rate $\kappa_g/\kappa_b$ for different values of $K$ and $N_{\mathrm{in}}$, over the operating range for a 3 dB drop in peak gain. The simulation results confirm our theoretical analysis: when the system is biased near the bright-eigenmode point, a reduction in $K$ at a fixed $N_{\mathrm{in}}$ improves the signal gain, but narrows the operating range of $\kappa_g/\kappa_b$. Conversely, for a fixed nonlinearity $K$, an increase in $N_{\mathrm{in}}$ reduces the signal gain but widens the operating range.

\emph{Noise suppression.—}Having established the signal amplification of the quadrature $\langle\mathbb{X}^{a}_{\mathrm{out}}\rangle$, we now analyze the quantum fluctuations to reveal the key signature of phase-sensitive noise suppression, originating from noise added to the conjugate output.

To this end, we capture the dynamics of quantum fluctuations using linearized QLEs, neglecting higher-order terms. The validity of this linearized model is discussed in the Supplemental Material \cite{sup_mat}. We define the quadrature fluctuation operators as
$\delta\!{{\mathbb{X}}_{\alpha}}=(\delta\!{\alpha^{\dagger}}e^{i\theta}+\delta\!{\alpha}e^{-i\theta})/\sqrt{2}$
and $\delta\!{{\mathbb{Y}}_{\alpha}}=i(\delta\!{\alpha^{\dagger}}e^{i\theta}-\delta\!{\alpha}e^{-i\theta})/\sqrt{2}$.
Their time evolution is governed by
\begin{equation}
\frac{d\bm{u}}{dt}=\mathcal{R}\bm{u}+\bm{\sigma},
\end{equation}
where $\bm{u}=$($\delta\!{\mathbb{X}_{a}}$, $\delta\!{\mathbb{Y}_{a}}$, $\delta\!{\mathbb{X}_{b}}$, $\delta\!{\mathbb{Y}_{b}})^\mathrm{T}$ is the column vector of quantum fluctuations, and $\bm{\sigma}$ is the column vector of noise sources given by
\begin{align}
\bm{\sigma} = \big(
& \sqrt{2\kappa_{a}}\delta\!\mathbb{X}^{a}_{\mathrm{in}}+\sqrt{2\kappa_{g}}[\mathrm{cos}(2\theta)\delta\!\mathbb{X}^{g}_{\mathrm{in}}-\mathrm{sin}(2\theta)\delta\!\mathbb{Y}^{g}_{\mathrm{in}}], \nonumber \\
&\sqrt{2\kappa_{a}}\delta\!\mathbb{Y}^{a}_{\mathrm{in}}-\sqrt{2\kappa_{g}}[\mathrm{sin}(2\theta)\delta\!\mathbb{X}^{g}_{\mathrm{in}}+\mathrm{cos}(2\theta)\delta\!\mathbb{Y}^{g}_{\mathrm{in}}], \nonumber \\
&\sqrt{2\kappa_{b}}\delta\!\mathbb{X}^{b}_{\mathrm{in}}, \sqrt{2\kappa_{b}}\delta\!\mathbb{Y}^{b}_{\mathrm{in}}\big)^\mathrm{T}.
\end{align}
In this basis, the coefficient matrix takes the form
\begin{equation}
\mathcal{R}=\left(
\begin{matrix}
\kappa_{n}+2\widetilde{K}_{y}& \Delta_{a}^{\prime}-2\widetilde{K}_{x}& 0&  J& \\
-\Delta_{a}^{\prime}-2\widetilde{K}_{x}& \kappa_{n}-2\widetilde{K}_{y}&  -J&  0&   \\
0& J& -\kappa_{b}& \Delta_{b}&  \\
-J& 0& -\Delta_{b}& -\kappa_{b}&  \\
\end{matrix}
\right),
\end{equation}
where $\Delta^{\prime}=\Delta+4K|\langle{a}\rangle|^{2}$ is the effective detuning induced by the Kerr nonlinearity, and $\widetilde{K}_{x}$ and $\widetilde{K}_{y}$ are the real and imaginary parts of $\widetilde{K}=K\langle{a}\rangle^{2}e^{-i2\theta}$, respectively.

\begin{figure}
\centerline {\includegraphics[width=8cm]{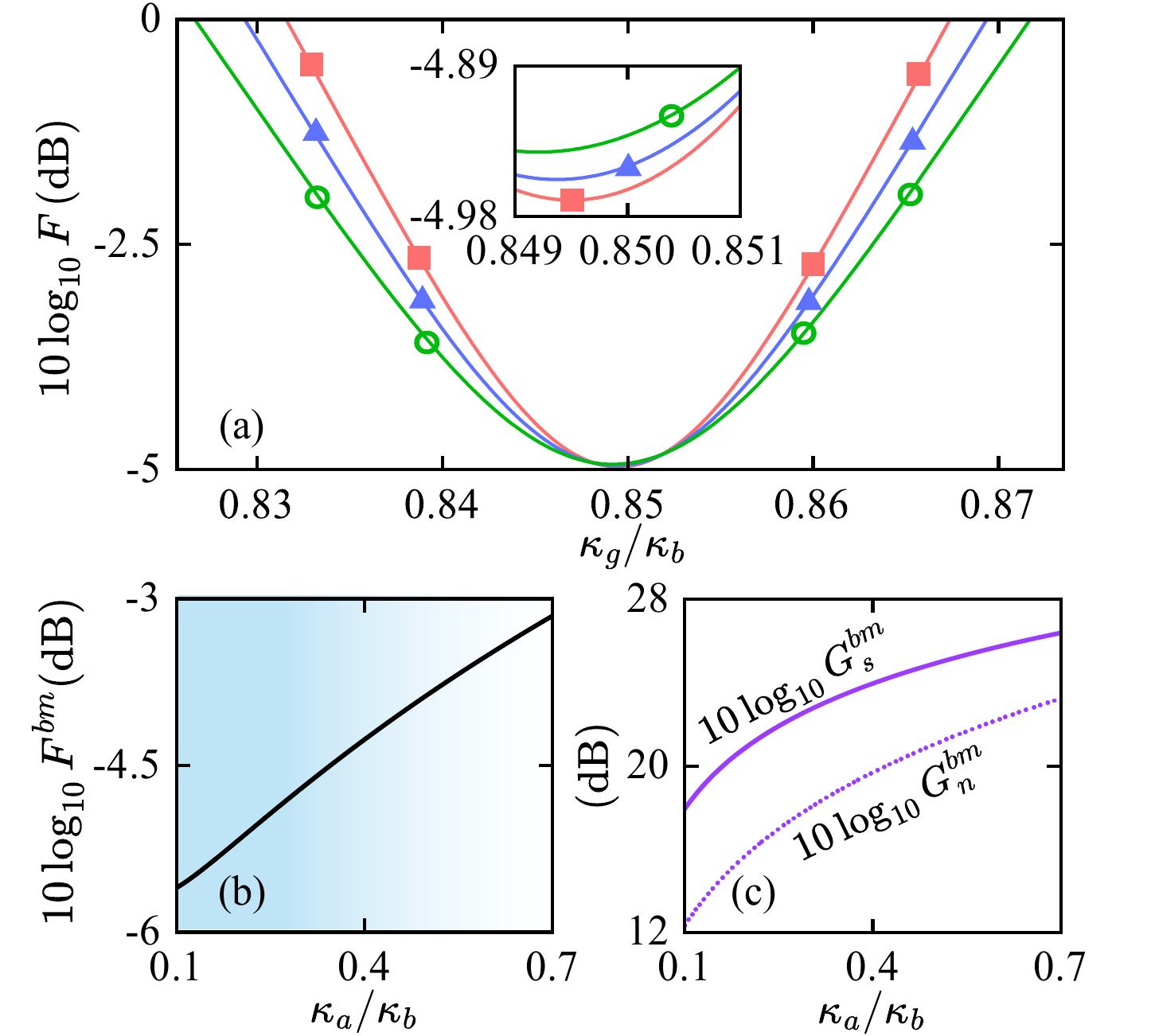}}
\caption{(a) Noise figure $10\mathrm{log_{10}}F$ versus normalized gain rate $\kappa_g/\kappa_b$
for: $N_\mathrm{in}=0.5$, $K=1\times10^{-4}\kappa_b$ (circle);
$N_\mathrm{in}=0.5$, $K=5\times10^{-5}\kappa_b$ (square); $N_\mathrm{in}=0.7$, $K=5\times10^{-5}\kappa_b$ (triangle).
(b) Noise figure $10\mathrm{log_{10}}F^{bm}$, and (c) signal gain $10\mathrm{log_{10}}G_{s}^{bm}$ (solid), noise gain $10\mathrm{log_{10}}G_{n}^{bm}$ (dotted) versus normalized decay rate $\kappa_a/\kappa_b$
for $N_\mathrm{in}=0.5$, $K=1\times10^{-4}\kappa_b$, $\kappa_n=0.6\kappa_b$.
Other parameters are the same as in Fig.~2.}
\end{figure}

The noise suppression is reflected in the output noise spectrum of the selected quadrature, defined as $\delta\!\mathbb{X}^{a}_{\mathrm{out}}=(\delta\!{a}_{\mathrm{out}}^{\dagger}e^{i\theta}+\delta\!{a}_{\mathrm{out}}e^{-i\theta})/\sqrt{2}$.
To derive the noise spectrum, we perform a Fourier transform of Eq.~(12) and apply the quantum input-output relation.
This yields the frequency-domain output quadrature in terms of the input noise operators:
${\delta\!}\mathbb{X}^{a}_{\mathrm{out}}(\omega)={\delta\!}\mathbb{X}_{\mathrm{in}}^{a}(\omega)
-\sqrt{2\kappa_{a}}\sum_{j=1}^{4}\mathcal {T}_{1j}(\omega){\sigma}_{j1}(\omega)$,
where $\mathcal{T}_{1j}(\omega)$ denotes the $(1,j)$ element of the matrix $\mathcal{T}(\omega)=-(\mathcal {R}+i\omega{\mathcal{M}_{0}})^{-1}$.
Then, we substitute this frequency-domain output quadrature into the output noise spectrum
$S_{\mathrm{out}}^{{\delta\!}\mathbb{X}_{a}}(\omega)
={\int_{-\infty}^{\infty}}\langle{{{\delta\!}\mathbb{X}_{\mathrm{out}}^{a}(\omega){\delta\!}\mathbb{X}_{\mathrm{out}}^{a}(\omega^{\prime})}
+{\delta\!}\mathbb{X}_{\mathrm{out}}^{a}(\omega^{\prime}){\delta\!}\mathbb{X}_{\mathrm{out}}^{a}(\omega)}\rangle{d\omega^{\prime}}/{4\pi}$. At resonance with the driving frequency ($\omega=0$), the noise spectrum becomes $\widetilde{S}_{\mathrm{out}}^{\delta\!\mathbb{X}_{a}}=G_{n}/2$.
Here, the factor of $1/2$ results from the vacuum noise contribution, and the resultant noise gain $G_{n}$ is given by
\begin{align} \nonumber
G_{n}=&|1-2\kappa_{a}\widetilde{\mathcal{T}}_{11}|^2
+4\kappa_{a}^2|\widetilde{\mathcal{T}}_{12}|^2+4\kappa_{a}\kappa_{g}(|\widetilde{\mathcal{T}}_{11}|^2+|\widetilde{\mathcal{T}}_{12}|^2) \\
&+4\kappa_{a}\kappa_{b}(|\widetilde{\mathcal{T}}_{13}|^2+|\widetilde{\mathcal{T}}_{14}|^2),
\end{align}
where full expressions of $\widetilde{\mathcal{T}}_{1j}$ are provided in the Supplementary Material \cite{sup_mat}.

\begin{figure}[tbph]
\centerline {\includegraphics[width=8cm]{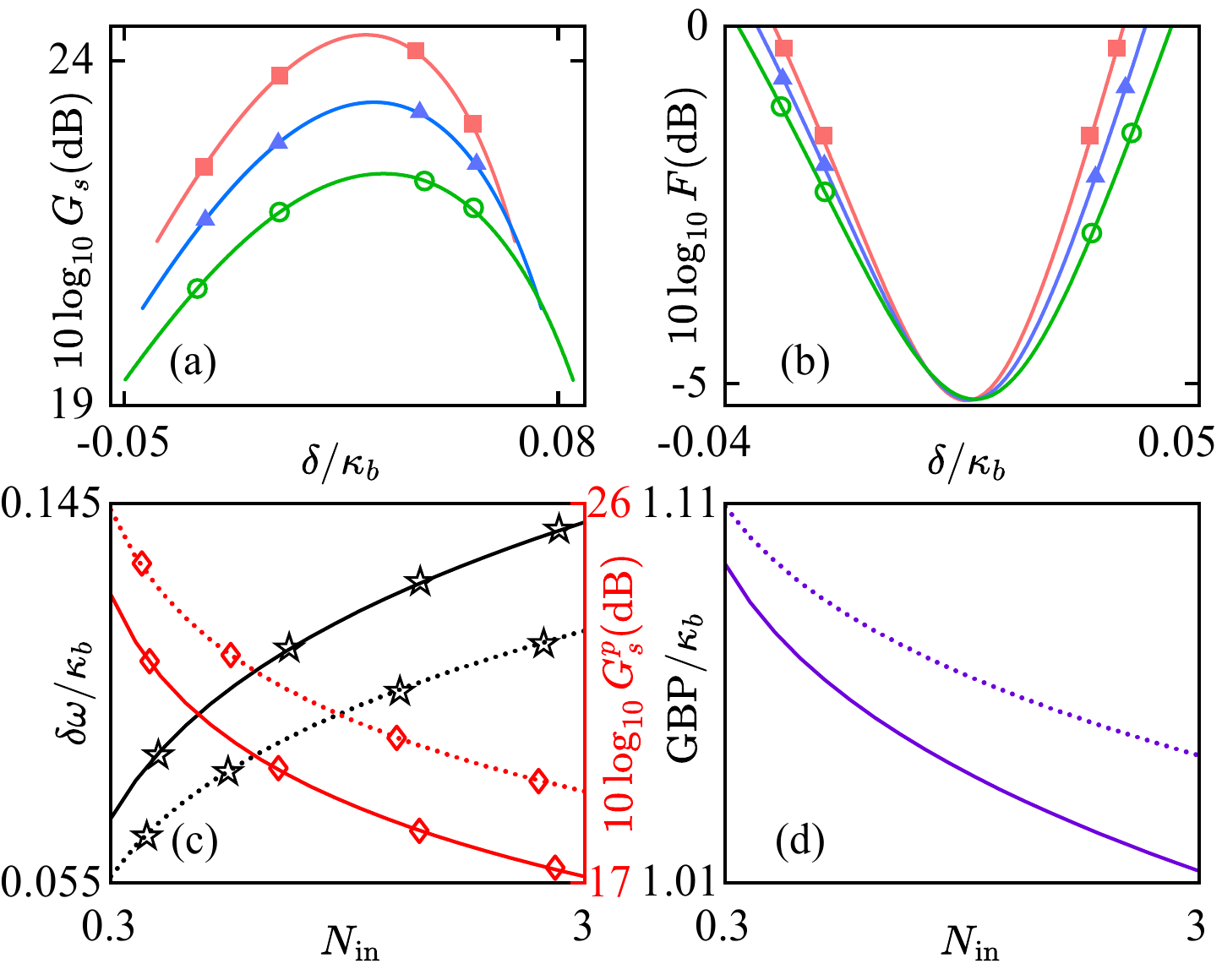}}
\caption{(a) Signal gain $10\mathrm{log_{10}}G_{s}$ and (b) noise figure $10\mathrm{log_{10}}F$ versus normalized detuning $\delta/\kappa_b$ for: $N_\mathrm{in}=0.5$, $K=1\times10^{-4}\kappa_b$ (circle);
$N_\mathrm{in}=0.5$, $K=5\times10^{-5}\kappa_b$ (square); $N_\mathrm{in}=0.7$, $K=5\times10^{-5}\kappa_b$ (triangle).
(c) Normalized operational bandwidth $\delta\omega/\kappa_b$ (star)
and peak signal gain $10\mathrm{log_{10}}G_{s}^{p}$ (rhombus) versus $N_\mathrm{in}$.
(d) Normalized $\mathrm{GBP}/\kappa_b$ versus $N_\mathrm{in}$.
In (c,d), $K=1\times10^{-4}\kappa_b$ (solid) and $K=5\times10^{-5}\kappa_b$ (dotted).
Parameters: $\omega_d=\omega_a-0.3\kappa_b+\delta$, $\kappa_n=0.6\kappa_b$; other parameters are the same as in Fig.~2.}
\end{figure}

We quantify this effective suppression using the noise figure $F$, defined as
\begin{align}
F=\frac{G_{n}}{G_{s}},
\end{align}
where $F<1$ implies the SNR enhancement. This regime is {\itshape unattainable} in ideal phase-sensitive linear quantum amplifiers without non-classical inputs, as they amplify signal and noise equally.
In stark contrast, our scheme exploits the Kerr nonlinearity to reshape both the signal and the quantum noise dynamics, thereby providing phase-sensitive noise suppression; that is, the noise in one quadrature is reduced while additional noise is added to its conjugate \cite{sup_mat}.

Figure~3(a) plots the noise figure $10\mathrm{log_{10}}F$ versus the normalized gain rate $\kappa_{g}/\kappa_{b}$ for different $K$ and $N_\mathrm{in}$. This confirms the predicted noise suppression by a clear drop below 0 dB around the bright-eigenmode point. Notably, at this operating point, the noise figure is analytically expressed as
\begin{align}
F^{bm}=\frac{1}{G_{s}^{bm}}+\frac{2(\kappa_a +\kappa_n)}{9\kappa_n},
\end{align}
which follows from the noise gain $G_{n}^{bm}=1+2G_{s}^{bm}(\kappa_{a}+\kappa_{n})/9\kappa_{n}$
and the signal gain $G_{s}^{bm}=\kappa_a(\kappa_{n}/N_{\mathrm{in}}^{2}\kappa_b^{2}K^2)^{1/3}$.

In the high-gain limit ($G_{s}^{bm}\gg1$), the noise figure reduces to $F^{bm}\approx2(\kappa_{a}+\kappa_{n})/9\kappa_{n}$. This simplified expression shows that for a fixed net gain rate $\kappa_n$, $F^{bm}$ depends solely on $\kappa_a$ and becomes independent of both $K$ and $N_{\mathrm{in}}$. The resulting pure $\kappa_a$-dependence of $10\log_{10}F^{bm}$ is shown in Fig.~3(b). Consequently, reducing $\kappa_a$ lowers the noise figure and enhances the noise suppression at the cost of reduced signal gain [see Fig.~3(c)]. Therefore, $\kappa_a$ fundamentally governs the trade-off between the noise suppression and the signal amplification.

Furthermore, although intrinsic losses in modes $a$ and $b$ are always present, the noise figure remains highly robust against such dissipation, as demonstrated in the Supplementary Material \cite{sup_mat}.

\emph{Operational bandwidth.—}Finally, we characterize the operational bandwidth $\delta\omega$ of our amplification scheme by driving the bright eigenmode with a detuned input field ($\omega_{d}\rightarrow\omega_{d}+\delta$). The bandwidth is taken as the frequency range satisfying two concurrent conditions: the signal gain $10\mathrm{log_{10}}G_{s}$ stays within 3 dB of its peak, and the noise figure $10\mathrm{log_{10}}F$ remains below 0 dB, as shown in Figs.~4(a,b), respectively. An intrinsic trade-off is observed in our results: while increasing the Kerr nonlinearity $K$ or the input mean excitation number $N_\mathrm{in}$ broadens $\delta\omega$, it inevitably suppresses the signal gain. This inverse relationship is directly illustrated in Fig.~4(c) via the opposing trends of normalized bandwidth $\delta\omega/\kappa_b$ and peak signal gain $10\mathrm{log_{10}}G_{s}^{p}$ versus $N_\mathrm{in}$ for different $K$.

To quantify the net outcome of this competition, we employ the gain-bandwidth product, $\mathrm{GBP}=\sqrt{G_{s}^{p}}\cdot\delta\omega$. Figure~4(d) shows that the normalized $\mathrm{GBP}/\kappa_b$ gradually decreases with increasing $K$ and $N_{\mathrm{in}}$, revealing that signal gain reduction outweighs bandwidth expansion.

\emph{Conclusion and discussion.—}In summary, we have proposed a nonlinear quantum amplifier that beats the fundamental quantum noise limits of linear amplifiers and achieves a net improvement in the SNR. This is accomplished via {\itshape strong phase-sensitive nonlinear amplification}, which leverages the cooperative effect between a gain-stabilized bright eigenmode and Kerr nonlinearity, all while eliminating the need for post-selection or non-classical resources. Our approach thereby establishes a new paradigm for amplifiers that simultaneously amplify signals and suppress noise, with broad potential to impact quantum technologies.

Moreover, this approach is directly applicable to existing experimental platforms. For instance, one can use two coupled superconducting microwave cavities, with Kerr nonlinearity provided by an embedded Josephson junction or superconducting qubit \cite{Boissonneault2010,Nigg2012,He2023,Zoepfl2023,Zapata2024}, and gain introduced via a negative resistance or conductance element \cite{Jiang2011,Xiao2019,Yao2023,Wang2025}. Similarly, an implementation could consist of two coupled optical cavities, where a $\chi^{(3)}$ nonlinear medium and rare-earth dopants supply the Kerr nonlinearity \cite{Kippenberg2004,MarinPalomo2017,Jang2019,Ghalanos2020,Otabe2024,Moille2025,Tritschler2025} and optical gain \cite{Peng2014,Chang2014,Miri2019,Oezdemir2019,Kuang2023}, respectively.

\emph{Acknowledgement.—}This work was supported by the National Natural Science Foundation of China
(Grants No. W2411002, No. 12375018, No. 12474363, No. 12505030 and No. 12574299),
the Natural Science Foundation of Shaanxi Province (Grant No. 31222000080029),
and the Innovation Program for Quantum Science and Technology (Grant No. 2021ZD0302001).
F. N. is supported in part by the Japan Science and Technology Agency (JST) [via the CREST Quantum Frontiers program Grant No. JPMJCR24I2,
the Quantum Leap Flagship Program (Q-LEAP), and the Moonshot R\&D Grant No. JPMJMS2061].

\bibliography{RefPR}

\end{document}